\documentclass[sigconf]{acmart}

\AtBeginDocument{%
  \providecommand\BibTeX{{%
    \normalfont B\kern-0.5em{\scshape i\kern-0.25em b}\kern-0.8em\TeX}}}

\usepackage[T1]{fontenc}
\usepackage[utf8]{inputenc}

\usepackage{graphicx}
\usepackage{url}
\usepackage{comment}
\usepackage{pgfplots,tikz}
\usepackage{graphics}
\usepackage{multirow,longtable}
\usepackage{xr}
\usepackage{xcolor}
\usepackage{amsmath}

\usepackage{amssymb}
\usepackage{algorithm}
\usepackage[noend]{algpseudocode}
\usepackage{eurosym}
\usepackage{ulem}

\definecolor{AAGreen}{rgb}{0.0, 0.5, 0.0}

\usepackage{tabto}

\newcommand\marginsymbol[1][0pt]{%
\color{#1}\tabto*{0cm}\makebox[\dimexpr-2pt\relax][r]{$\bullet$}\tabto*{\TabPrevPos}}
\newcommand{\rem}[3]{
    \marginsymbol[#1]
	\color{#1}\footnote{\textbf{\textcolor{#1}{#2}}:  \textit{#3}
	\hrule} \color{black}
}
\newcommand{\modif}[2]{\color{#1}#2 \color{black}}

\newcommand{\modifNT}[1]{\modif{blue}{#1}}

\newcommand{\remJ}[1]{\rem{red}{Jihane}{#1}}
\newcommand{\modifJ}[1]{\modif{red}{#1}}

\newcommand{\modifAA}[1]{\modif{AAGreen}{#1}}

\newcommand{\MDG}[0]{ModelDrivenGuide}
\newcommand{\anonymous}[1]{[ANONYMOUS]}

\makeatletter
\newcommand{\costVec}[2][r]{%
  \gdef\@VORNE{1}
  \left(\hskip-\arraycolsep%
    \begin{array}{#1}\vekSp@lten{#2}\end{array}%
  \hskip-\arraycolsep\right)}

\def\vekSp@lten#1{\xvekSp@lten#1;vekL@stLine;}
\def\vekL@stLine{vekL@stLine}
\def\xvekSp@lten#1;{\def\temp{#1}%
  \ifx\temp\vekL@stLine
  \else
    \ifnum\@VORNE=1\gdef\@VORNE{0}
    \else\@arraycr\fi%
    #1%
    \expandafter\xvekSp@lten
  \fi}
\makeatother

\pgfplotsset{compat=1.17}

\begin{document}

\title{How to Optimize the Environmental Impact of Transformed NoSQL Schemas through a Multidimensional Cost Model?}









 \author{Jihane Mali}
 \affiliation{%
   \institution{ISEP}
   \city{Paris}
   \country{France}
 }
 \email{jihane.mali@isep.fr}

 \author{Faten Atigui}
 \affiliation{%
   \institution{CEDRIC, Conservatoire National des Arts et M\'etiers (CNAM)}
   \city{Paris}
   \country{France}
}
 \email{faten.atigui@cnam.fr}

 \author{Ahmed Azough}
 \affiliation{%
   \institution{L\'eonard de Vinci P\^ole Universitaire, Research Center}
   \city{Paris La D\'efense}
   \country{France}
 }
 \email{ahmed.azough@devinci.fr}

 \author{Nicolas Travers}
 \orcid{0000-0002-3502-151X}
\affiliation{%
   \institution{L\'eonard de Vinci P\^ole Universitaire, Research Center}
   \city{Paris La D\'efense}
   \country{France}
 }
 \email{nicolas.travers@devinci.fr}

 \author{Shohreh Ahvar}
 \affiliation{%
   \institution{Nokia Networks}
   \city{Massy}
   \country{France}
 }
 \email{shohreh.ahvar@nokia.com}

\renewcommand{\shortauthors}{Mali, et al.}

\begin{abstract}
The complexity of database systems has increased significantly along with the continuous growth of data, resulting in NoSQL systems and forcing Information Systems (IS) architects to constantly adapt their data models (i.e., the data structure of information stored in the database) and carefully choose the best option(s) for storing and managing data. 
In this context, we propose 
an automatic global approach for leading data models' transformation process. This approach starts with the generation of all possible solutions. It then relies on a cost model that helps to compare these generated data models in a logical level to finally choose the best one for the given use case.
This cost model integrates both data model and queries cost.
It also takes into consideration the environmental impact of a data model as well as its financial and its time costs.
This work presents for the first time a multidimensional cost model encompassing time, environmental and financial constraints, which compares data models leading to the choice of the optimal one for a given use case. 
In addition, a simulation 
for data model's transformation and cost computation has been developed based on our approach.
\end{abstract}


 


\maketitle

\section{Introduction}
\label{IntroductionContxt}

Data's explosion especially characterized  by the \texttt{3V} (Volume, Variety \& Velocity) has opened up major research issues related to modeling, manipulating, and storing massive amounts of data ~\cite{tauro2012}.

The resulting so-called NoSQL systems correspond to \textit{four families} of data structures: key-value oriented (\texttt{KVO}), wide-column oriented (\texttt{CO}), document oriented (\texttt{DO}), and graph oriented (\texttt{GO}). These systems have raised new problems of transforming traditional databases into non-relational databases, whether in terms of storage management, data query, cost or performance~\cite{chebotko2015,li2014,vajk2013,rocha2015,daniel2016umltographdb,abdelhedi2017,claudino2016,DELAVEGA2020455}.
Thus, this highly distributed context emphasizes the problem of integrating an Information System (IS) and finding the proper data model is essential for such systems.
Furthermore, environmental protection regulations are constantly evolving, financial competition between different NoSQL solutions is increasing, and time efficiency is a constant challenge.
Requirements and needs are constantly growing thus, guaranteeing the efficiency and availability of information requires the ability to respond effectively to new demands and requests on information systems. Better restructuring of database schemas is needed to maintain the 3V promise.

The main issue is to provide the optimal data model for a given Information System usage. To solve this issue, we  need the cost estimation of a solution according to the data model, statistics and queries to compute~\cite{ozsu1999}. This issue is hardly tackled for NoSQL solutions and especially when choosing the target architecture and structure.


To tackle this issue, we previously proposed a data model transformation approach \textit{ModelDrivenGuide}~\cite{Mali2020,Mali2022} 
which aims at proposing 
a set of data models providing choices instead of focusing on a dedicated solution which prevents any trade-off and optimizing only one cost dimension.
Each data model is transformed from previous data models to offer the possibility to structure data in a more suitable manner.
Then \textit{ModelDrivenGuide} reduces the search space by taking into account the use case (a set of queries). 
Budgets and latency have always been important factors to guide the choice of a data model. In addition, in respect to climate change awareness and moving towards a responsible consumption for SDG~12\footnote{UN SDG 12: \url{https://www.un.org/sustainabledevelopment/sustainable-consumption-production/}}, the issue of respecting environmental impact arose. 
In this work, we propose a cost model to compare the generated data models and to lead the choice of the best data model(s). 
This cost model is essential since it avoids implementing all possible solutions (thousands of data models~\cite{Mali2022}), and measuring each cost to choose the best one.
The cost model is of different dimensions, time, environmental and financial, and consists of both query independent cost (\textit{i.e.,} data model) and query dependent cost (\textit{i.e.,} all use cases queries on a data model). It depends on multiple parameters to define the cost of each data model (\textit{e.g.,} queries, data volume, number of servers, etc.).
This cost remains at logical level in order to compare data models efficiency and not NoSQL solutions themselves.


In this paper, our main contributions are: 
\begin{itemize}
    \item A multidimensional cost model, which integrates three key dimensions: time, environmental and financial costs,
    \item Advanced simulations and evaluations of environmental cost to compare impacts of data models while compared to competitors' solution.
\end{itemize}

The rest of the paper is organized as follows: a state of art is presented in Section,~\ref{sec:related works}, a context for data model denormalization in Section~\ref{sec:DM generation}. Our multidimensional cost model is defined in Section~\ref{sec:cost model}. 



\section{Related Works}
\label{sec:related works}



The main contribution of this work 
is to propose a cost model that aims to estimate the cost of a data model, so that we can decide whether this data model is the most suitable for our use case among  all the possible ones. To do so, we propose a multidimensional cost model that considers both query execution time, financial cost, and environmental impact.
In this regard, existing work have focused either on data models transformation or on measuring query costs, or the environmental impact in the context of optimization.



\paragraph{Data Model Transformation}


Many previous studies have addressed the issue of transforming an existing data 
into a new one more suitable to handle growing data. These studies focus on transforming data models to align with one or multiple NoSQL families. Some approaches specifically convert a relational data model into a target NoSQL data model, such as \texttt{CO}~\cite{vajk2013}, \texttt{DO}~\cite{rocha2015} or \texttt{GO}~\cite{daniel2016umltographdb}.
Additionally, other studies opt to transform their data models into one or multiple physical solutions. Chebotko et al.~\cite{chebotko2015} introduce a query-driven approach for modeling \textit{Cassandra} DB starting from an ER model. They define dedicated transformation rules between these logical and physical models. Similarly, De Freitas et al.~\cite{claudino2016} propose a conceptual mapping approach that converts ER models into one of the four NoSQL families, utilizing an abstract formalization of the mapping rules.

Other studies employ MDA to transform UML class diagrams into NoSQL databases. Li Y. et al. ~\cite{li2014} propose a transformation approach that converts a class diagram into a \texttt{CO} database using \textit{HBase}, while Daniel G. et al.~\cite{daniel2016umltographdb} transform it into a \texttt{GO} database.
Abdelhedi et al.~\cite{abdelhedi2017} suggest transforming a class diagram into a unified logical model that encompasses the four families of NoSQL databases. Subsequently, this logical model is further transformed into physical models specific to each NoSQL family.
In addition, Alfonso et al. present the system \textit{Mortadelo}~\cite{DELAVEGA2020455}, which introduces a model-driven design process capable of generating implementations for \texttt{CO} and \texttt{DO} databases based on the conceptual model.

\begin{figure*}[ht]
     \centering
     \includegraphics[width=\linewidth]{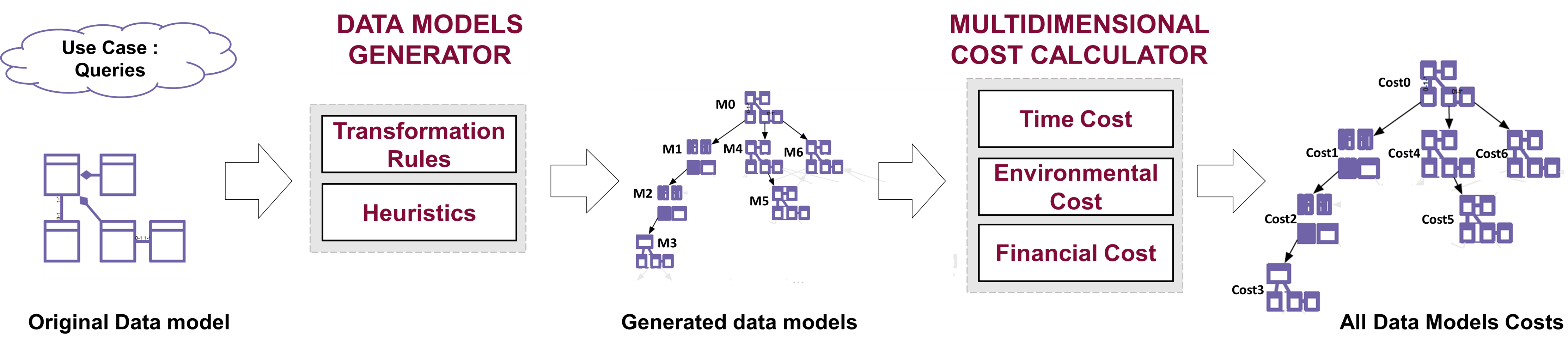}
     \caption{\MDG{}: Global Model Driven Approach}
     \label{fig:MDG_Process}
 \end{figure*}

\paragraph{Query Cost Model}

In the literature, cost models have mainly been used in the context of query optimization,  
some existing work have considered the time to execute a query to compare execution plans. 
In~\cite{ozsu1999}, the authors express the cost of a distributed execution strategy with respect to either the total time or the response time. The total time is the sum of all time components (also referred to as cost), while the response time is the elapsed time from the initiation to the completion of the query. The cost includes the local processing time (time of a CPU instruction and the time of a disk~I/O) and the communication time (the fixed time to initiate and receive a message, and the time it takes to transmit a data unit from one site to another).
In~\cite{sellami2017}, the authors define the cost of an execution plan as the sum of the cost of each operation (\textit{i.e.,} node in the execution plan) composing an execution plan. This cost depends at least on one of the three costs that are the CPU cost, the input/output cost, and the communication cost.

\paragraph{Environmental Cost of IS} 

Other research papers focus on examining the environmental impact with carbon footprints. The substantial data processing in cloud servers consumes a significant amount of energy, resulting in substantial carbon emissions and environmental consequences.
In~\cite{thakur2016carbon}, the authors explore various technologies aimed at reducing energy consumption and carbon footprints in data centers. They also establish metrics for quantifying carbon emissions.
In Rodriguez et al.~\cite{rodriguez2011}, the authors develop a cost model that utilizes multilinear regression to estimate the global power and energy cost associated with database queries. Instead of measuring power/energy at the individual hardware component level (such as CPU, RAM, HD), the cost models are derived from two sources: a) measurements obtained from internal sensors or a power meter attached to the server enclosure, and b) readily available workload statistics like relation cardinality, tuple size, number of columns, and number of servers. In Saraiva et al.~\cite{saraiva2017}, the authors focus on evaluating the energy consumption of different query types while comparing relational and NoSQL database approaches, specifically MySQL and Neo4j.
Mahajan et al.~\cite{mahajan2019} investigate query optimization techniques that enhance energy efficiency without compromising the performance of both relational and NoSQL databases.

\paragraph{Discussion} 
All the studies conducted for model transformations aim to formalize and/or automate the rules for transforming a source model into a target model. However, none of them have specifically focused on encompassing all possible models or providing guidance for selecting a model based on well-defined criteria. To the best of our knowledge, our work is the first to propose such an approach that facilitates the identification of all potential solutions for a given use case, utilizing a multidimensional cost model.

The studies carried out in the context of query optimization mainly propose cost models to assess the costs of query execution. None of these works consider measuring the different costs of queries according to the different data models.

Regarding the environmental cost, most of the literature's works primarily focus on measuring the energy consumption of queries during their execution. However, none of them have proposed  an environmental cost specifically for a data model that encompasses all relevant parameters, including the data models themselves and the specific use case. 

As far as we know, our work is the first to introduce a cost model specifically designed for a NoSQL data model. This cost model considers both the structure of the data model (independent of queries) and the queries that are primarily performed on it.
Our goal is to provide a comprehensive cost model to compare data models with each other, and a global view to guide the process of selecting an appropriate data model that ultimately minimizes query costs. We propose to consider the time cost, the environmental impact, and the financial cost.


\section{Context: Data Models Generation}
\label{sec:DM generation}

Our approach \texttt{\MDG{}} aims to generate logical data models (Definition~\ref{def:DM}) for each family (NoSQL and relational). It provides a modeling framework based on these logical models and transformation rules (\textbf{Merge} in Definition~\ref{def:merge} and \textbf{Split} in Definition~\ref{def:split}).

 \begin{figure*}[t]
\begin{minipage}[t]{.36\linewidth}
	\centering\includegraphics[width=.9\linewidth]{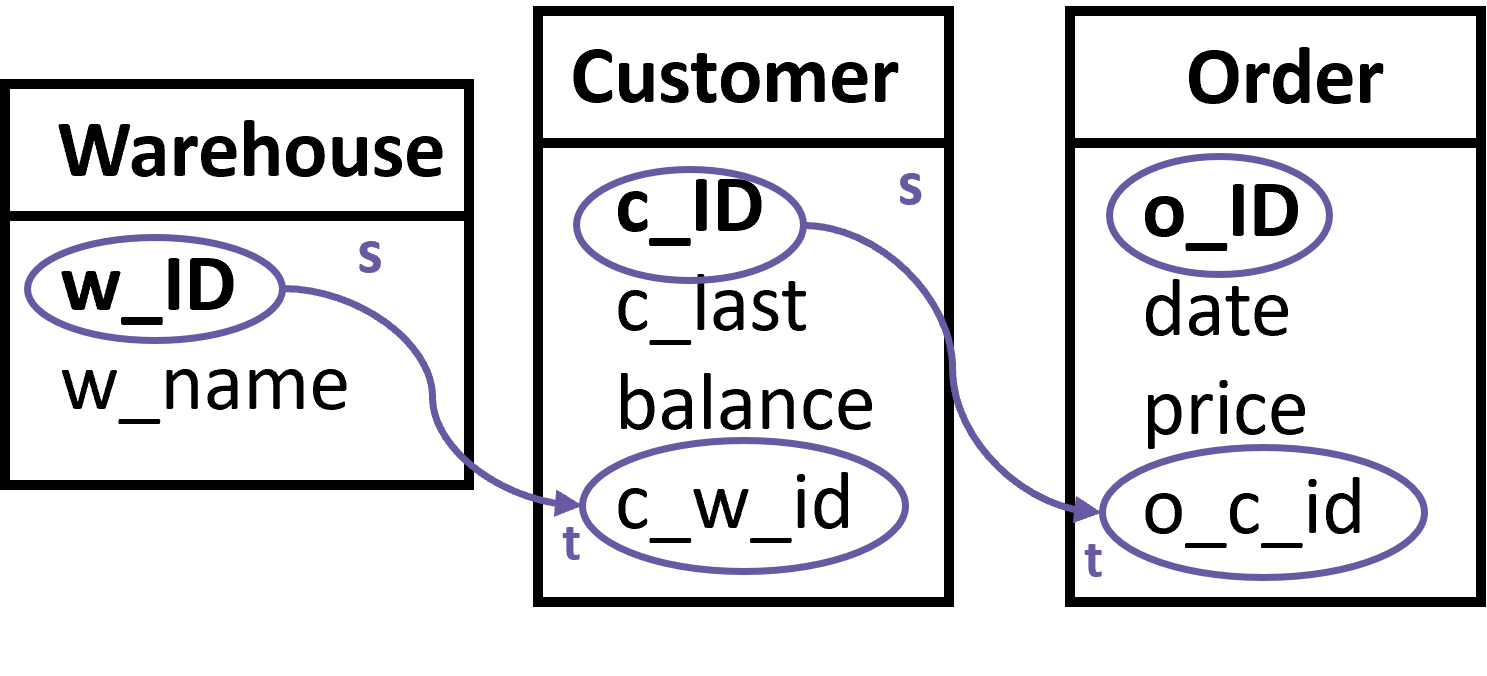}
	\caption{Data Model Example ($M0$\label{fig:data model example})}
\end{minipage}
\hfill
\begin{minipage}[t]{.24\linewidth}
	\centering\includegraphics[width=.99\linewidth]{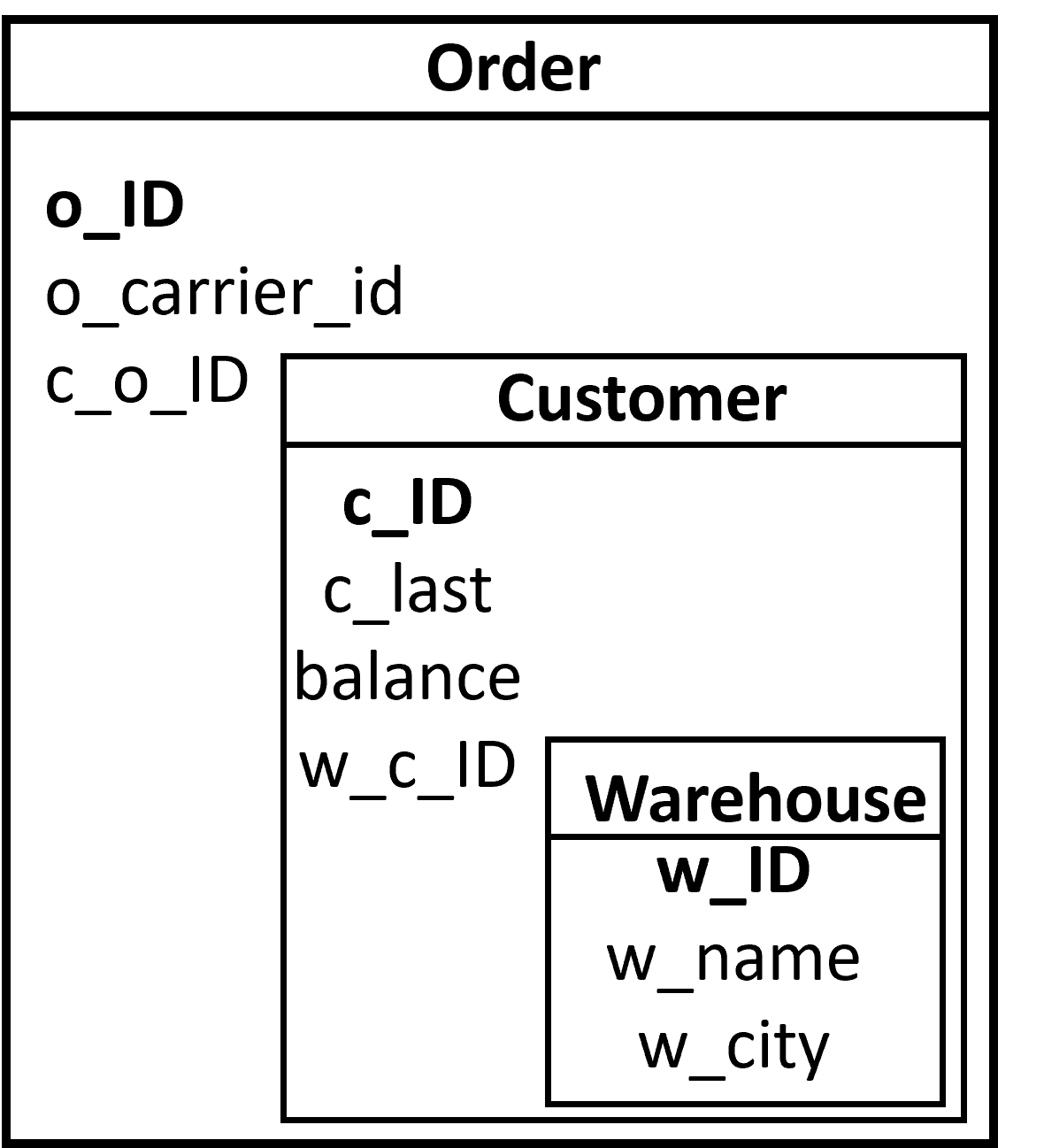}
	\caption{Merge Example ($M35$) \label{fig:merge}}
\end{minipage}
\hfill
\begin{minipage}[t]{.38\linewidth}
	\centering\includegraphics[width=.99\linewidth]{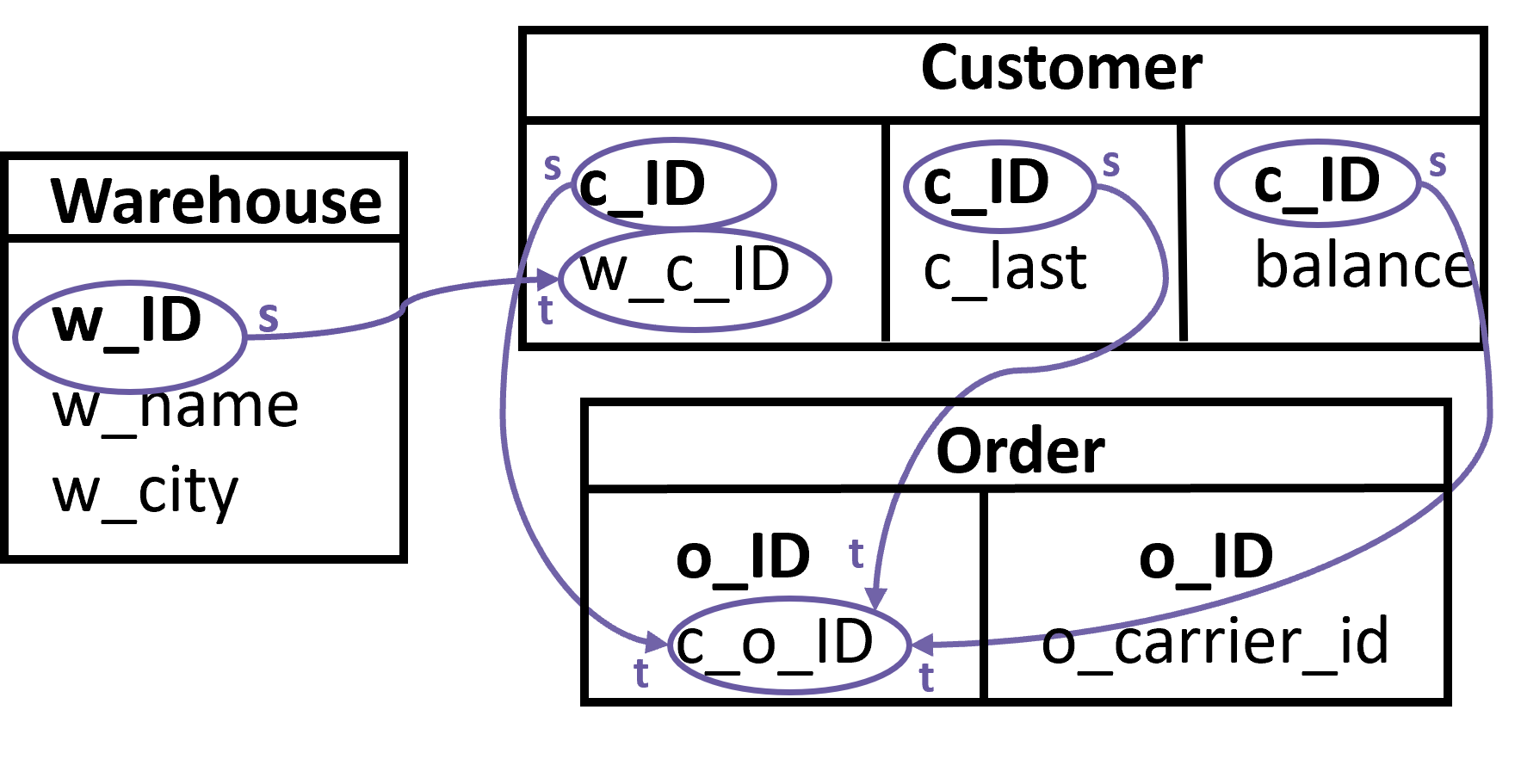}
	\caption{Split Example ($M3$) \label{fig:split}}
\end{minipage}
\end{figure*}

\begin{definition}{}\label{def:DM} Let ${\mathcal M}$ be a data model conform to the  meta-model 5Families~\cite{Mali2020} where ${\mathcal M} = ({\mathcal C}, {\mathcal R}, {\mathcal L}, {\mathcal K}, {\mathcal E}, \kappa)$ is composed of 
\texttt{concepts} $c(r_1,...,r_m) \in {\mathcal C} | r_1,...,r_m\in {\mathcal R}$, 
\texttt{rows} $r(k_1,...,k_n) \in {\mathcal R} | k_1,...,k_n \in {\mathcal K}$, 
\texttt{key values} ${\mathcal K}$ (Atomic Values or Complex Values), 
\texttt{references} $ref_{i\rightarrow j} \in {\mathcal L} | k_i, k_j \in {\mathcal K}$, 
\texttt{edges} ${\mathcal E}: {\mathcal C} \times {\mathcal C}$ and 
\texttt{constraints} $cons(k) \in \kappa | k \in {\mathcal K}$.
\end{definition}

\texttt{\MDG{}} starts from the conceptual model, then goes from one logical data model ${\mathcal M}$ to another by applying refinement rules recursively. 
This generation process allows generating \textbf{all} possible data models and thus, allows choosing the targeted solution. However, the number of possibilities explodes as splits on ${\mathcal M}$ can be applied on each key and merges can be done in both ways.
 
To tackle this complexity issue, we proposed a 
heuristic that allows reducing the search space and avoid cycles. It is based on the idea of avoiding to produce the same data models with two different paths. In fact, applying splits and merges in different orders will produce the same effects on the resulting data models. Moreover, every merges can be reversed by a split and produce a cycle in the production of data models.


\begin{definition}{}\label{def:merge}
Let \textit{m}: ${\mathcal M}\rightarrow {\mathcal M}$ be an endogenous function that merges rows from a model ${\mathcal M}$. The merge function $m(r_i,r_j,\-ref_{i\rightarrow j})$ is applied on two rows $r_i, r_j \in {\mathcal R}$ linked by a reference $ref_{i\rightarrow j} \in {\mathcal L}$ where $r_i$ and $r_j$ are the source and target rows with corresponding keys $k_i, k_j \in {\mathcal K}$.
The merge function produces a new model ${\mathcal M}'$ where $r_j$ is a complex value of $r_i$, and removes $ref_{i\rightarrow j}$, denoted by: $$m(r_i,r_j,ref_{i\rightarrow j}) = r_i\{r_j\}$$

The merge function is a bijective function $m^{-1}(r_i\{r_j\}) = (r_i,r_j,\-ref_{i\rightarrow j})$
which rebuilds $ref_{i\rightarrow j}$ and non\-nested rows.
\end{definition}
 
\begin{definition}{}\label{def:split}
Let \textit{s}: ${\mathcal M}\rightarrow {\mathcal M}$ be an endogenous function that splits rows from a 5Families model ${\mathcal M}$. The split function $s(r_i,k)$ is applied on a row $r_i \in {\mathcal R}$ and a key value $k \in keys(r_i)$ not linked to a constraint.
The split function produces a new model ${\mathcal M}'$ with two rows $\overline{r_{i_{k}}}$ and $r_{i_k}$ with the same constraint key $pk \in keys(r_i)$ (\textit{i.e.,} primary key) where $\overline{r_{i_{k}}} = (pk, k_i)|\forall k_i \in keys(r_i) \land k_i \neq k$, and $r_{i_k} = (pk, k)$.
The split function $s$ is denoted by: 
$$s(r_i,k) = (\overline{r_{i_{k}}},r_{i_k})$$

$s$ is a bijective function $s^{-1}(\overline{r_{i_{k}}},r_{i_k}) = (r_i)$
which merges common constraints and keys.
\end{definition}


In the following, denormalization of data models will refer to the combination of merge and split operations.
Notice that the cost model manipulates \textit{rows} as defined in Definition~\ref{def:DM} (a schema as a set of keys from a concept) and not instances (we denote them by \textit{documents}).

\paragraph{\textbf{Driving Example:}} Let be a normalized data model $M0$ in Figure~\ref{fig:data model example} with 3 rows, \textit{Warehouse}, \textit{Customer}, and \textit{Order} (resp. $W$, $C$ and~$O$).
All denormalized data models are produced by recursively splitting and/or merging keys and rows from this normalized data model. 

Figure~\ref{fig:merge} shows an example of a denormalized data model generated by applying two merge functions on data model $M0$. It consists of 1 row $O$, containing the nested row $C$ that, in turn, contains the nested row $W$.

Furthermore, Figure~\ref{fig:split} depicts an example of a denormalized data model generated by applying split operations recursively on $M0$. It consists of 6 rows $W$, $C1$, $C2$, $C3$, $O1$, and $O2$. We notice two recursive splits were applied on the row $C$ which resulted in three different rows where the primary key (\texttt{c\_ID}) was duplicated and the rest of keys were split. As for the row $O$, one split was applied to generate two rows $O1$ and $O2$.

We must notice that \MDG{} proposes 36 different data models~\cite{Mali2022} among thousands of possible ones. Those 36 data models target at least one query from the use case.
\section{Multidimensional Cost Model}
\label{sec:cost model}

Our data model's generation process allows to generate a set of data models (Left part of Figure~\ref{fig:MDG_Process}). 
Besides conventional measures like response time and throughput, NoSQL database systems demand higher requirements due to the massive amount of data they need to handle (which incurs significant costs) in terms of storage, processing and communication.

In order to choose the best data model out of a set of generated possible solutions ${\mathcal M}$, we propose a cost model (Definition~\ref{def:cost}) that automatically calculates the costs of logical data models to compare them and choose the best one.
The main contribution of this work is to propose a multidimensional cost model based on  \textbf{time} $T$, \textbf{environmental} $E$ and \textbf{financial} $F$ cost functions (as shown in Figure~\ref{fig:MDG_Process} - central part).

       

\begin{definition}{}\label{def:cost}
Let $M \in {\mathcal M}$ a data model and ${\mathcal Q}$ = \{$q_1$,...,$q_n$\} be a set of queries from the use case. The multidimensional cost function ${\mathcal C}$   
of $M$ regarding ${\mathcal Q}$ is defined by:
$${\mathcal C}(M,{\mathcal Q}) = \costVec{T(M,Q);E(M,Q);F(M,Q)} =\costVec{T(M);E(M);F(M)}
+ \sum_{{i=1}}^{n} \omega_{q_i} \times \costVec{T(M, q_i);E(M, q_i);F(M, q_i)}$$

\noindent where $\phi \in \{T,E,F\}$ are cost functions on the data model $M$ either independent ($T(M), E(M), F(M)$) or dependent ($T(M, q_i), E(M, q_i), \-F(M, q_i)$) on queries $q_i \in {\mathcal Q}$ , with their related query's average daily occurrences 
$\omega_{q_i}$ 
$T, E, F$ are sub-functions corresponding respectively to the time, environmental and financial dimensions of the cost model.
\end{definition}


To achieve this we need to measure these subfunctions with common parameters: volumes and servers.
Each cost dimension relies on the volume of stored data, the volume of processed data on servers and the volume of transferred data among servers.
Moreover, each cost dimension combines those volumes in different ways, depending mostly on the data model itself and queries computation.

\paragraph{Volume of Storage}
Each data model has its own data size since some redundancy occurs through merges (rows' duplication) and splits (primary keys' repetition). 
Thus, the volume of storage is calculated for each data model.
This volume has an impact on processed data that needs to be read on the servers, and on the number of servers required to store the whole database.


\paragraph{Volume of Processing}
The computation of queries requires to read data either on SSD drives (for cold starts) or main memory (RAM).
Notice that the CPU time is negligible in this context. Hence the CPU environmental impact is encompassed into the RAM impact.
    
\paragraph{Volume of Communication}
It assesses the cost of moving or transferring data between different servers. 
A distinction has to be made between inter and intra-datacenter communications, particularly in terms of financial cost, since service providers charge fees for external communications.


\medskip
Table~\ref{tab:constants} shows the constants and variables involved in the multidimensional cost model described throughout this article.
\begin{table*}[t]

\begin{minipage}[t]{.58\linewidth}
\centering
\small
\begin{tabular}{|l|p{1cm}|r|r|r|}
\hline
& \textbf{Variable} & \textbf{Time} & \textbf{Environmental} & \textbf{Fees}\\
&& \textbf{constant~$C^T_{x}$} & \textbf{ impact~$C^E_{x}$} &  $C^F_x$\\
\hline
\hline
Data processed in RAM & $V_{RAM}$ & 1.25 GB/s & 0.0280  kg CO$_2$e/GB & \\
\hline
Data stored in SSD & $V_{SSD}$ & 0.325 GB/s & 0.0031  kg CO$_2$e/GB&\\
\hline
\multirow{2}*{Bandwidth for data transfers} & $V_{COM}^{ext}$ & 1.0 GB/s & 0.0110  kg CO$_2$e/GB & 0.019 \euro/GB\\
 & $V_{COM}^{int}$ & 1.0 GB/s & 0.0110  kg CO$_2$e/GB &\\
\hline
1 server & $\#srv$ & N/A & 0.87671 kg CO$_2$e/day & 0.8543\euro/day\\
\hline
\end{tabular}
\caption{Cost Models variables \& Constants\label{tab:constants}}
\end{minipage}
\hfill
\begin{minipage}[t]{.32\linewidth}
    \centering
    \small
    \begin{tabular}{|l|l|}
    \hline
     \textbf{Notation} & \textbf{Description}\\ 
     \hline\hline
     $M$ & a Data Model among solutions ${\mathcal M}$\\
     \hline
     ${\mathcal R}$ & Row from ${\mathcal M}$ composed of keys  \\
     \hline
     ${\mathcal L}$ & Links between rows  \\
     \hline
     ${\mathcal K}$ & Key (composed of nested rows)  \\
     \hline
     ${\mathcal Q}$ & Query  \\
     \hline
     $doc$ & Entity (Object) of a row  \\
     \hline
     $\pi_q(doc)$ & Query result set with projection $\pi$ \\
     \hline
     $sel_k$ & Selectivity of query's filter key $\sigma_k$ \\
     \hline
    \end{tabular}
    \caption{Cost Models Notations}
    \label{tab:my_label}
\end{minipage}
\end{table*}

\subsection{Time Cost Dimension}
\label{seq:timecost}

The time cost of a data model can vary according to several factors, including the size and complexity of the data model, the used storage type and processing infrastructure, and the speed of the network connection.


\begin{definition}\label{def:timecost}
Let $T(M,q)$ be the time cost of a query $q \in \mathcal {Q}$ on a data model $M \in \mathcal{M}$ that evaluates the time expressed in seconds required for the query $q$ to process and transmit the data. Let $T(M)$ be the independent time cost of the data model $M$. We denote:

\begin{eqnarray}\label{eq:T}
T(M, q) = \frac{V_{RAM}^{T}(M,q)}{C_{RAM}^{T}} + \frac{V_{SSD}(M,q)}{C_{SSD}^{T}} + \frac{V_{COM}(M,q)}{C_{COM}^{T}}\\ 
T(M) &=& 0 \nonumber
\end{eqnarray}


\end{definition}

$T(M,q)$ is calculated based on the volume of data processed by RAM access $V_{RAM}^{T}$, storage access on SSD $V_{SDD}$, as well as the volume of data transmitted $V_{COM}$ in Bytes.
Constants speed like $C_{RAM}^{T}$, $C_{SSD}^{T}$ and $C_{COM}^{T}$ are expressed in GB/s (GigaByte per second) and correspond to standard values.

Notice that $T(M) = 0$ since the computation time is only dependent on queries.

\subsection{Environmental Cost Dimension}
\label{seq:envcost}

Processing queries has an impact on the energy consumption. In fact, the environmental cost depends on data accesses like RAM, storage and communication. 
The cost model needs to quantify the amount of data processed for each query according to each of the data models in order to compare them.

Measuring the exact environmental impact is an impossible task, especially as most studies focus on the global impact of systems\footnote{\url{https://medium.com/teads-engineering/evaluating-the-carbon-footprint-of-a-software-platform-hosted-in-the-cloud-e716e14e060c}} rather than the details of individual treatments. Our aim is therefore to estimate consumption for the purposes of comparing data models, and not to obtain a precise impact.

Environmental impact constants depend on several parameters such as the data center's energy source, the server manufacturer, the recycling process, etc.
All the constants presented in Table~\ref{tab:constants} (environment and fees) correspond to an \textit{Azure B1s}\footnote{\url{https://learn.microsoft.com/en-us/azure/virtual-machines/sizes-b-series-burstable}} instance in France. 
We have set  the environmental impact constants based on various studies on network~\cite{guyon2018}, SSD I/O~\cite{tannu2022dirty} and RAM\footnote{RAM impact: \url{https://ismaelvelasco.dev/emissions-in-1gb}}.
All constants can be changed (region, provider, RAM, SSD, Comm).



Notice that the environmental footprint of a server is independent of queries $E(M)$ where servers whole lifecycles are studied~\cite{10.1109/MM.2022.3163226}. We can consider that, on average, a single server corresponds to 320 kg of CO2e/year (0.87671 kg CO2e/day).
In our case, $E(M)$ is influenced by the number of servers required by a data model. Indeed, some data models contain more redundancy and require more RAM and disk space, hence the number of servers required.


\begin{definition}\label{def:environmentalcost}
Let $E(M,q)$ be the environmental cost of a query $q \in \mathcal {Q}$ on a data model $M \in \mathcal{M}$  that returns the carbon footprint of $q$ while processing, storing and transmitting data. Let $E(M)$ be the independent environmental cost of the data model $M$. We denote:
\begin{eqnarray}\label{eq:E}
E(M, q)& =& V_{RAM}^{E}(M,q) \times C_{RAM}^{E} + V_{SSD}(M,q) \times C_{SSD}^{E}\\
 &&+ V_{COM}(M,q) \times C_{COM}^{E}\nonumber\\
E(M) &=& \#srv \times C_{srv}^{E}\nonumber
\end{eqnarray}
\noindent where $E(M) \& E(M,q)$ are expressed in kg CO2e.
\end{definition}

Notice that the volume of data read in main memory considered in $V_{RAM}^{E}$ is different from $V_{RAM}^T$ (Section~\ref{sec:query}) while volumes for communications $V_{COM}$ and storage $V_{SSD}$ are identical.
Constants are expressed in kg CO2e/GB: $C_{RAM}^{E}$, $C_{SSD}^{E}$ and $C_{COM}^{E}$.

\subsection{Financial Cost Dimension}
\label{seq:fincost}

The financial cost of a data model has few dependency on query execution, since most of the expenses come from the number of servers $\#srv$ depending on the pricing model (\textit{e.g.,} pay-as-you-go or subscription).
However, most service providers have fees for data transfers outside of the datacenter\footnote{\url{https://azure.microsoft.com/en-us/pricing/details/bandwidth/}}
which means that our cost model needs to distinguish internal from external communications.

\begin{definition}\label{def:financialcost}
    Let $\mathcal{\textit{F}(M)}$ be the financial cost of a query $q \in \mathcal {Q}$ on a data model $M \in \mathcal{M}$  that returns the financial cost of a data model $M$. Let $F(M)$ be the independent financial cost of the data model $M$. We denote:
\begin{eqnarray}\label{eq:F}
F(M,q)&=&V_{COM}^{ext}(M,q) \times C_{COM}^F\\
F(M) &=& \#srv \times  C_{srv}^{F}\nonumber
\end{eqnarray}
\noindent where $F(M)$ and $F(M,q)$ are expressed in currency (\textit{e.g.,} \euro, \$).
\end{definition}

\subsection{Data Model's Cost}

Our cost model relies on the computation of data volumes in RAM, SSDs and networks. For this, we need to detail the main operations applied to the database in order to estimate the queries' cost. In the following, we detail filtering and join queries using traditional algorithms. The computation varies depending on sharding keys and local indices. 

Note that denormalization also has an impact on joins (more with splits, less with merges), but also on data reads from RAM (and vice versa).
Our approach attempts to find the best tradeoff among data model transformations.


\subsubsection{Filter queries ($\sigma_k$)}
\label{sec:query}
The filter operation applies a selection $\sigma$ on database instances according to a given filtering key $k\in q$.
The cost computation depends on the number of targeted servers (\#srv).
Thus, the functions of volume $V_{RAM}$ and $V_{COM}$ in our cost model express the combination of each server cost.



First, a query is sent to the servers (its number can vary) and results are given back to the application.
The amount of data transfer (Equation~\ref{equ:COMvar}) has an impact on the global bandwidth; thus, we apply a sum of all communications $v_{COM}^{s_i}$ (Equations~\ref{equ:shardCaseCOM} \& \ref{equ:IndexCaseCOM}).
\begin{equation}\label{equ:COMvar}
   V_{COM} = \sum_{{i=1}}^{\#srv} v_{COM}^{s_i}
\end{equation}

Considering the computation time on servers $V^T_{RAM}$, thanks to parallelism, each server $s_i$ runs independently of the others.
The total time (Equation~\ref{equ:RAMtimeVar}) to process data on all servers is the maximum processing time on a single server $v_{RAM}^{s_i}$ (Equations~\ref{equ:shardCaseRAM} \& \ref{equ:IndexCaseRAM}).
\begin{equation}\label{equ:RAMtimeVar}
V_{RAM}^{T} = \max(v_{RAM}^{s_i}), \forall i \in \{1,2,...,\#srv\}
\end{equation}



On the other hand, the environmental impact (Equation~\ref{equ:RAMenvVar}) of a filter query takes into account every single data processing on all the servers. It is then impacted by the sum of volume processed on each server $v_{RAM}^{s_i}$ (Equations~\ref{equ:shardCaseRAM} \& \ref{equ:IndexCaseRAM}).
\begin{equation}\label{equ:RAMenvVar}
   V_{RAM}^{E} = \sum_{{i=1}}^{\#srv} v_{RAM}^{s_i}
\end{equation}

It is important to notice for read queries that $V_{SSD} = 0$ for warm start (best cache hit ratio~\cite{10.1145/3959.3961}). Moreover, we can simplify $V_{SSD} = V_{RAM}$ for update queries.

\medskip
Volume functions $V$ rely on local costs  $v^{s_i}$ on the set of servers $\mathcal{S} = \{s_1,s_2,...,s_n\}$. The number of processed data depends on the data placement strategy, sharding and indices :



\paragraph{a) Sharding-based filters}
When a filter query $\sigma_k$ implies the sharding key~$k$, data are available on a single server.
Thus, the number of data processed on this server corresponds to the number of corresponding documents to access the sharded key $|shard_k|$ and to read it in main memory (doc size $|doc|$ \& selectivity $sel_k$).
Otherwise, there will be no processing on other servers (Equation~\ref{equ:shardCaseRAM}).
\begin{equation}\label{equ:shardCaseRAM}
    v_{RAM}^{s_i} = \begin{cases}
        |shard_k| + |doc| \times \#doc \times sel_k, & \text{on server } s_i,\\
        0, & \text{otherwise.}
    \end{cases}
\end{equation}

Recall for documents' size, thanks to data model denormalizations, implies volume changes $|doc|$ (nesting \& splits).


According to the volume, a data transferred for each server $v_{COM}^{s_i}$, it must take into account the query size $|q|$ and the volume of output documents projected on required keys $|\pi_q(doc)|$ (Equation~\ref{equ:shardCaseCOM}). 
\begin{equation}\label{equ:shardCaseCOM}
    v_{COM}^{s_i} = \begin{cases}
        |q| + |\pi_q(doc)| \times \#doc \times sel_k, & \text{on server } s_i,\\
            0, & \text{otherwise.}
    \end{cases}
\end{equation}

\paragraph{b) Index-based filters}
When no sharding is available for a filter key $k$, all servers need to process the query even if no corresponding data is available.
Of course, the selection requires a scan in the related index $|index_k|$ instead of the whole local data.

In order to estimate the number of servers that could answer, the filtering key $\#srv_k$ we combine the number of documents and $k$'s selectivity (Equation~\ref{equ:nbrSrvs}).
\begin{align}\label{equ:nbrSrvs}
    \#srv_k = \begin{cases}\begin{array}{lr}
        1, & \text{if } \exists shard_k,\\
        \min(\lceil \#doc \times sel_k \rceil, \#srv), & \text{otherwise.}
        \end{array}
        \end{cases}
\end{align}

Thus, the number of data read on each server relies on an index scan and data access when available on $\#srv_k$ servers  (Equation~\ref{equ:IndexCaseRAM}).
\begin{equation}\label{equ:IndexCaseRAM}
    v_{RAM}^{s_i} = \begin{cases}
            |index_k| + |doc| \times \left\lceil \frac{\#doc \times sel_k}{\#srv_k} \right\rceil,  & \text{if } i \in \{1,2,...,\#srv_k\},\\
            |index_k|, & \text{otherwise}.
            \end{cases}
\end{equation}

Finally, the volume of data transferred by each server depends on query's size (to query all servers), projected documents' size $\pi_q$ and the number of documents provided locally by the query (Equation~\ref{equ:IndexCaseCOM}).
\begin{equation}\label{equ:IndexCaseCOM}
    v_{COM}^{s_i} = \begin{cases}
            |q| + |\pi_q(doc)| \times \left\lceil \frac{\#doc \times sel_k}{\#srv_k} \right\rceil,  & \text{if } i \in \{1,2,...,\#srv_k\},\\
            |q|, & \text{otherwise}.
            \end{cases}
\end{equation}

\paragraph{c) No index filters}
The worst case occurs when no sharding nor index are available for a filter query $\sigma_k$.
It requires to process all data on each server (Equation~\ref{equ:NoIndexCaseRAM}). But for communications, Equation~\ref{equ:IndexCaseCOM} remains applicable.
\begin{equation}\label{equ:NoIndexCaseRAM}
    v_{RAM}^{s_i} = \frac{|doc| \times \#doc}{\#srv}, \forall i \in \{1,2,...,\#srv\} 
\end{equation}






\subsubsection{Join Queries}
To process join queries on a distributed environment, some works propose various MapReduce~\cite{10.5555/1251254.1251264} based implementations~\cite{10.1145/1739041.1739056,10.1145/1807167.1807273,ALBADARNEH20221074}. However, NoSQL databases avoid this heavy process by applying denormalization~\cite{chebotko2015,abdelhedi2017,sellami2017}. Moreover, at this level of abstraction we put off addressing the precise implementation question as we are comparing data models and not NoSQL solutions.

However, since our approach compares data models costs, we need to apply a join process for those for which there is no merge between the two required rows.
For this, we apply a simple strategy with a nested loop join where the order is driven by queries' selectivity.
Its cost is integrated in Algorithm~\ref{algo:query cost} by combining all sub queries.

\begin{algorithm}[t]
\caption{Query's Cost}\label{algo:query cost}
\begin{algorithmic}[1]
\Statex \textbf{global: Data model $M \in \mathcal{M}$, query $q \in \mathcal{Q}$, $rows \in M$}
\Statex \textbf{input: $keys \in q$} 
\Statex \textbf{init: ${\mathcal C} =0, cost = 0, \#output=0$} 
    
    \Procedure{QueryCost}{$keys$}
        \State $rows \gets getCoveredRows(keys)$
        \ForAll{$r \in rows$}
            \State $rowKeys = q \cap r$
            \State $cost\gets \costVec{T(r, rowKeys);E(r, rowKeys);F(r, rowKeys)}$
            \Comment{Equations~\ref{eq:T}, \ref{eq:E} \& \ref{eq:F}}
            \State $nb \gets sel_{rowKeys} \times \#doc_r$             
            \If{${\mathcal C} =0$}
                \State ${\mathcal C}=cost$
                \State $\#output \gets nb$
            \Else
                \State ${\mathcal C} \gets {\mathcal C} + \#output \times cost$
                \State $\#output \gets \#output \times nb$
            \EndIf
        \EndFor
    \EndProcedure
\end{algorithmic}
\end{algorithm}

\subsubsection{Query Cost Computation}

To compute the cost of a given data model, we need to combine the previous formulas according to the data model structure and the keys involved in the queries.
To achieve this, we propose the following algorithm, which builds an execution plan in the form of iterations on the rows to be processed for a given query and data model.

Algorithm~\ref{algo:query cost} constitutes the main process for computing query cost.
It checks operations (\textit{e.g.,} joins, filters, etc.) to be performed on rows $rows \in M$ required by a query $q$.

First, we must get the list of $rows \in M$ implied in the query wrt. to the $keys \in q$. The function $getCoveredRows$ (line~2) extracts the list of rows which minimizes its size (a key can occur in several rows with denormalization) to avoid unnecessary joins by covering $q$.
Moreover, this list is ordered according to the selectivity of required join operations to reduce the cost of this step.

Then, for each row (line~3), we extract keys $rowKeys$ implied by query $q$ in current row $r$ (line~4).
We then perform operations on $r$ and calculate its cost $cost$ on different dimensions (line~5) following the previously defined equations~\ref{eq:T}, \ref{eq:E} \& \ref{eq:F}.
After applying the operation, we estimate the number of the produced instances $nb$ (line~6) depending on the filter's selectivity. This value will be used to compute the join cost.


For the first queried row (lines~7-9) we initialize the cost and the number of outputs.
In other cases (\textit{i.e.,} join queries), we apply the nested loop join (line~11) by adding the costs to $\mathcal{C}$ and estimate the output size (line~12).
Iteratively we compute all required operations until all rows are being processed (line~3).

We must notice that the join algorithm presented in Alg~\ref{algo:query cost} constitutes a first step and alternative solutions can be used to compute the join cost.

\paragraph{\textbf{Driving Example:}} Let's take query $\mathcal{Q}4$ from Table~\ref{tab:queries} to illustrate Algorithm~\ref{algo:query cost} on the example data models presented previously.
$\mathcal{Q}4$ requires a join between \textit{Customer} and \textit{Order} rows, a filter operation on $c\_last$ and a projection of keys $o\_carrier\_id,c\_ID,o\_ID,c\_o\_ID$.

The first step of Algorithm~\ref{algo:query cost} is to get covered rows from a given query (line~2).
For data model $\mathcal{M}0$ (Figure~\ref{fig:data model example}), both rows \textit{Customer} and \textit{Order} are necessary to cover all these keys; thus, $||rows|| = 2$.
According to data model  $\mathcal{M}35$ (Figure~\ref{fig:merge}) requires a single row  $O$ since $C$ is nested into $O$; thus, $||rows|| = 1$.
Finally, for $\mathcal{M}3$ (Figure~\ref{fig:split}), the 3 rows $C2$ (for c\_last), $O1$ (for c\_o\_ID) and $O2$ (for o\_carrier\_id) are needed to cover all of query's keys, thus  $||rows|| = 3$.
This first step illustrates the fact that denormalization has an impact on query cost and provides different sequences of operation.

Then, the algorithm checks the row and keys implied by the query $q$ and computes the corresponding cost of the selected data model.
For the filter on row $O$ with key $c\_last$, the data model $M0$ produces an output set of documents (line~7-9) that must be joined with $C$ during the second loop (line~3).
While for data model $M35$, the filter on the nested document avoids the join since all keys are embedded in each single document.
But for data model $M3$, two joins are required to both rebuilding the split document between $C1$ and $C2$ and joining the row $O$.
Each loop takes into account the size of the output and filters selectivity to compute the cost of each operation.

Thus, the final cost for each data model can be resumed by:

\begin{eqnarray}
{\mathcal C}(M0,q) &=& \costVec{T(C, c\_last);E(C, c\_last);F(C, c\_last)} + \#output_C \times \costVec{T(O, o\_id);E(O, o\_id);F(O, o\_id)}\nonumber\\
{\mathcal C}(M35,q) &=& \costVec{T(O\{C\{W\}\}, c\_last\&o\_id);E(O\{C\{W\}\}, c\_last\&o\_id);F(O\{C\{W\}\}, c\_last\&o\_id)}\nonumber\\
        {\mathcal C}(M3,q) &=& \costVec{T(C2, c\_last);E(C2, c\_last);F(C2, c\_last)} + \#output_{C2} \times \costVec{T(O1, o\_c\_id);E(O1, o\_c\_id);F(O1, o\_c\_id)}\nonumber\\
        &&\hspace*{2.1cm}+ \#output_{O1} \times \costVec{T(O2, o\_id);E(O2, o\_id);F(O2, o\_id)}\nonumber
    \end{eqnarray}
    
M35 is not necessarily the best data model even if it does not require joins, in fact, it requires more memory to read long nested documents. The cost model helps to take into account all dimensions at the same time and make data models comparable.

\section{Implementation}
\label{sec:exp}

To compute cost models, we implemented in \textit{Java} a simulator which integrates the whole process presented in Figure~\ref{fig:MDG_Process}.
It starts by generating all data models as implemented in previous works.
It then calculates costs of each generated data model for a given setting couple $S \in \mathcal{S}, S =\{\#doc,\#srv\}$.

Our simulator starts by studying the structure of the generated data models based on a signature file which is the output of the data model's generation process.
Each data model has a unique signature that describes it by integrating the list of concepts, containing the list of corresponding rows with  keys and nested rows, and also the list of references linking rows (a sample of simplified signatures are represented in Table~\ref{tab:driving example}). 

Then, the simulator varies the settings couple (data volume \& servers) and updates elements of each data model accordingly (\textit{i.e.,} document size).
Finally, the simulator, based on our cost model, uses Equations~\ref{equ:COMvar}, \ref{equ:RAMtimeVar} \& \ref{equ:RAMenvVar} to define the volume of data processed, transferred and stored for each case and follows Algorithm~\ref{algo:query cost} to define the costs of queries on each data model as well as the total cost of a data model.

\begin{table*}[t]
\centering
\small
	\begin{tabular}{|l|l|l|p{2.5cm}|p{2.5cm}|l|r|r|}
		\hline
		\textbf{Query} & \textbf{Type} & \textbf{Filter Keys } & \textbf{Projection keys}& \textbf{Join keys}& \textbf{Sharding} &  \textbf{Occurrences} & \textbf{Constraints $\tau_{t_q}$}\\
            &&&&&&per day&(in s)\\
		\hline
		\textbf{$\mathcal{Q}1$}& Filter & $balance$ & $c\_last$ & N/A & N/A & 500 & $10^{-1}$\\
		\hline
		\textbf{$\mathcal{Q}2$}& Filter & $o\_ID$ & $o\_carrier\_id$ & N/A & N/A & 1,000 & $10^{1}$\\
		\hline
		\textbf{$\mathcal{Q}3$}& Filter & $w\_city$ & $w\_name$ & N/A & on $w\_city$ & 100 & $10^{0}$\\
		\hline
		\textbf{$\mathcal{Q}4$}& Join & $c\_last$ & $c\_last,o\_carrier\_id$ & $c\_ID,c\_o\_ID$ & on $c\_last$ & 50 & $10^{-1}$\\
		\hline
		\textbf{$\mathcal{Q}5$}& Join & $c\_last,w\_city$ & $balance,w\_name,$ $o\_carrier\_id$ & $c\_ID,c\_o\_ID,w\_ID,$ $w\_c\_ID$ & on $w\_city$ \& $c\_last$ & 10 & $10^{1}$\\
        \hline
	\end{tabular}
	\centering\caption{Use Case Queries\label{tab:queries}}
 \end{table*}
\begin{table}[t]
    \centering
    \small
    \begin{tabular}{|c|c|p{1.8cm}|}
		\hline
		\textbf{Data Model}& \textbf{Signature} & \textbf{Produced by competitor}\\
		\hline%
		\textbf{$M0$}& $W,C,O$&\\
		\hline
		\textbf{$M3$}& $W,C1,C2,C3,O1,O2$&\\
		\hline
		\textbf{$M16$}& $C1,C2,C3\{W,O\}$&\\
		\hline
        \textbf{$M24$}& $C1,C2\{W,O\}$&\\
		\hline
        \textbf{$M30$}& $W,C\{O\}$&\\
        \hline
        \textbf{$M33$}& $W,O\{C\}$&Chebotko~\cite{chebotko2015}\\
		\hline
        \textbf{$M35$}& $O\{C\{W\}\}$&Abdelhedi~\cite{abdelhedi2017}\\
		\hline
	\end{tabular}
	\caption{Generated Data Models and Signatures\label{tab:driving example}}
\end{table}

\begin{figure}[b]
	\centering\small
     \includegraphics[width=1.05\linewidth]{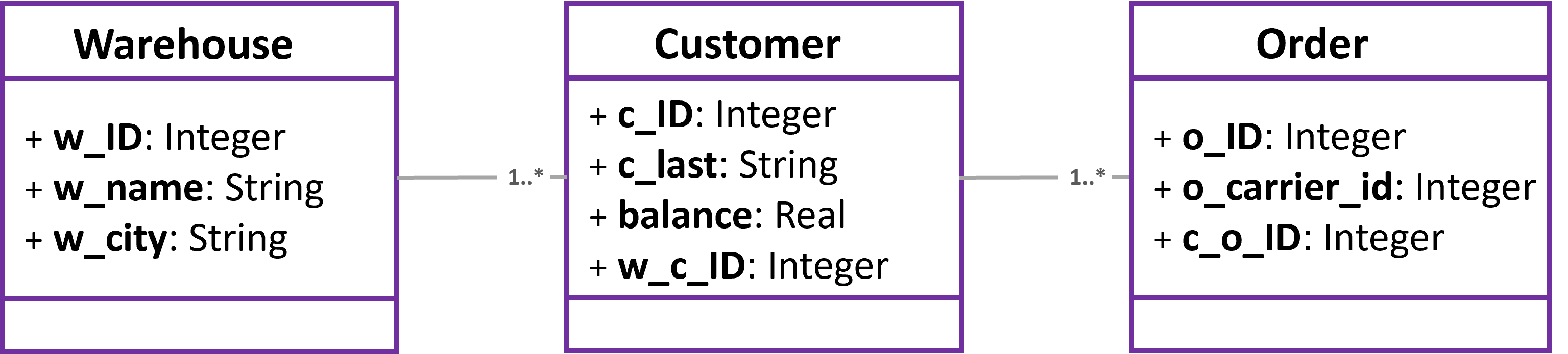}
     \caption{TPC-C Class Diagram}
     \label{fig:TPCC_class_diagram}
\end{figure}

\subsection{Use Case}

To illustrate our approach, we did a simulation using the TPC-C\footnote{ \url{http://www.tpc.org/tpc_documents_current_versions/pdf/tpc-c_v5.11.0.pdf}} benchmark giving a full use case with a set of queries mixing at the same time transactions, joins and aggregations. For this simulation, we focus on the three rows: \textit{Warehouse}, \textit{Customer}, and \textit{Order} depicted in Figure~\ref{fig:TPCC_class_diagram} .
In TPC-C, the number of documents of each row is factored by the number of warehouses. Therefore, in the following, we will refer to data volume by the number of warehouses.
For each warehouse, we have around 30,000 customers and 60,000 orders for an average of two orders per customer. 
Using our TPC-C example in Figure~\ref{fig:TPCC_class_diagram} as an initial data model $M0$ and use case queries listed in Table~\ref{tab:queries}  as inputs to our generation process, we have generated 36 data models by applying recursive denormalizations (3 splits and 32 merges).
Those data models along with the use case are the input to our cost model tool.

The 5 queries in Table~\ref{tab:queries} are segmented in filter, projection and join keys in order to show the different aspects of those queries, and especially operations applied on the database.
Filter keys impact costs on sharding (Equation~\ref{equ:shardCaseRAM} 
) and indices (Equations~\ref{equ:IndexCaseRAM} and~\ref{equ:IndexCaseCOM}) especially on selectivity.
While projections keys impact communication costs (Equation
and~\ref{equ:IndexCaseCOM}).
Finally join keys implies the number of loops in Algorithm~\ref{algo:query cost} (less for merged data models).

Queries $\mathcal{Q}1$, $\mathcal{Q}2$ and $\mathcal{Q}3$ are filter queries which target a row of the initial data model $M0$ (resp. $C$,$O$ and $W$).
Queries $\mathcal{Q}4$ and $\mathcal{Q}5$ are join queries that require a join between ($C$, $O$) and ($W$, $C$, $O$) respectively.
We varied query optimization strategies s.t. the first two queries do not use a sharding key.

Moreover, the complete use case contains also queries occurrences per day ($\omega_{q_i}$) to have a global impact of the database on the multidimensional cost.
It serves as a weight to the query which affects the total cost of a data model as specified in Definition~\ref{def:cost}. Table~\ref{tab:queries} gives each query's occurrences as defined in our configuration files.


\subsection{Denormalized Data Models}
In order to illustrate our approach, we have chosen 7 different data models out of the 36 generated ones that are representative of different classes of costs on queries.
Table~\ref{tab:driving example} shows the data models' signatures which define their structure:  rows separated by commas, nested rows with braces \{\}, split rows with numbers (\textit{e.g.,} C1, C2). Here are the selected data models:
\begin{itemize}
    \item $M0$ is the initial data model for data model's generation process (relational) and it consists of three rows:\\Warehouse $W$, Customer $C$ and Order $O$,
    \item $M3$ contains 6 rows, $C1$, $C2$ and $C3$ resulted from a two splits on $C$, $O1$ and $O2$ from a split on $O$ and $W$,
    \item $M16$ consists of 3 rows by applying a split~\ref{def:split} $C$ into three rows $C1$, $C2$ and $C3$ and merging~\ref{def:merge} $O$ and $W$ into $C3$,
    \item $M24$ consists of 2 rows by splitting $C$ in two rows $C1$, $C2$ and by nesting $O$ and $W$ in $C2$,
    \item $M30$ contains 2 rows $C$ resulted by nesting $O$ in $C$ and the second row is $W$,
    \item $M33$ produces the opposite merge between O and C (compared to $M30$). This data model is also produced when applying the \textit{Chebotko Diagram~\cite{chebotko2015}} one of our competitors dedicated to the Cassandra NoSQL solution,
    \item $M35$ contains a single row resulted by nesting $W$ in $C$ then $C$ in~$O$. This data model is obtained by Abdelhedi's approach~\cite{abdelhedi2017} when choosing the MongoDB target.
    
\end{itemize} 

To compare our global approach (\textit{i.e.,} choice of a data model among possible ones) with the related works, most strategies are query driven or model driven and provide a single output data model.
Query driven~\cite{chebotko2015}  relies on most frequent queries where ${\mathcal Q}1, {\mathcal Q}2$ and ${\mathcal Q}3$ target $C, O$ and $W$ independently but ${\mathcal Q}4$ will provide a merge led by $O$ (due to ${\mathcal Q}2$) thus, it produces ${\mathcal M}33$.
For MDA~\cite{abdelhedi2017,DELAVEGA2020455} a better compromise is given by applying splits and merges among all queries which produces $M35$.

It is interesting to notice that our competitors' solutions are available in our search space. Our cost model will compare their strategies and show which kind of optimization they provide.

Note that as defined in~\ref{def:cost}, data model's global cost $\mathcal{C}(M,\mathcal{Q})$ includes costs independent (\textit{e.g.,} storage cost, servers allocation cost) and costs dependent from queries.

\subsection{Data Model's Cost}

Our implemented cost model selects the optimal data model. 
It calculates the multidimensional cost of each data model in different parameters. The latter are integrated into the simulator through configuration files (\textit{e.g.,} data model's structure, setting variations, key sizes, indices, queries' information, etc.). Our simulator allows to study the impact of each data models' structure on queries' cost, and the impact of settings (\textit{i.e.,} data volume, number of servers) 
on data models costs. It then generates for each data model and setting, the costs of every query on this latter, as well as its total cost.

\begin{figure}
     \centering 
     \vspace*{-.5cm}
     \includegraphics[width=1.02\linewidth]{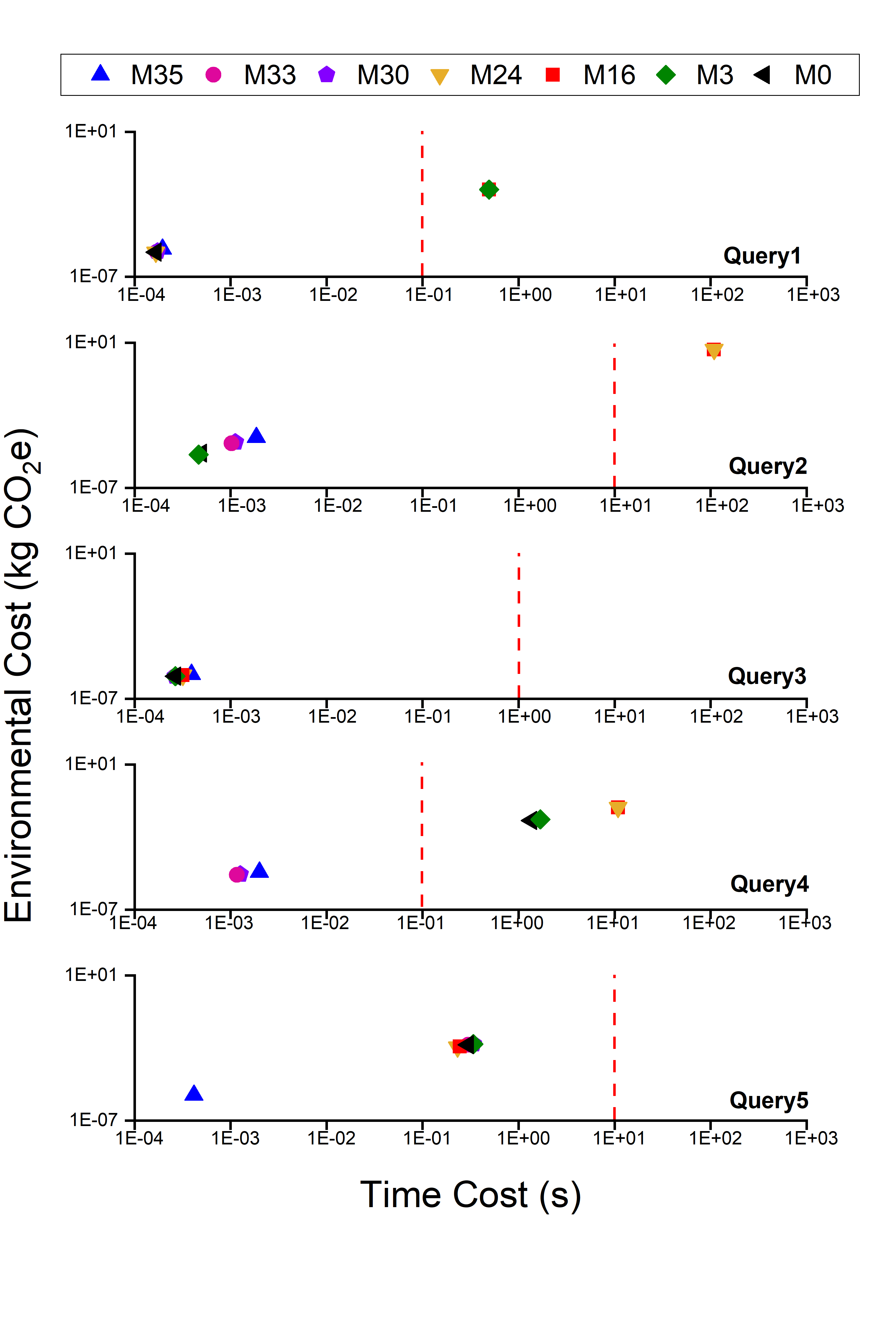}
     \caption{Unitary Query Cost per Data Model with: ($\#doc$ = 1M
     warehouses
     , $\#srv$ = 1,000)} \label{fig:queriescost}
\end{figure}
\begin{figure*}[t]
      \centering
         \includegraphics[width=\linewidth]{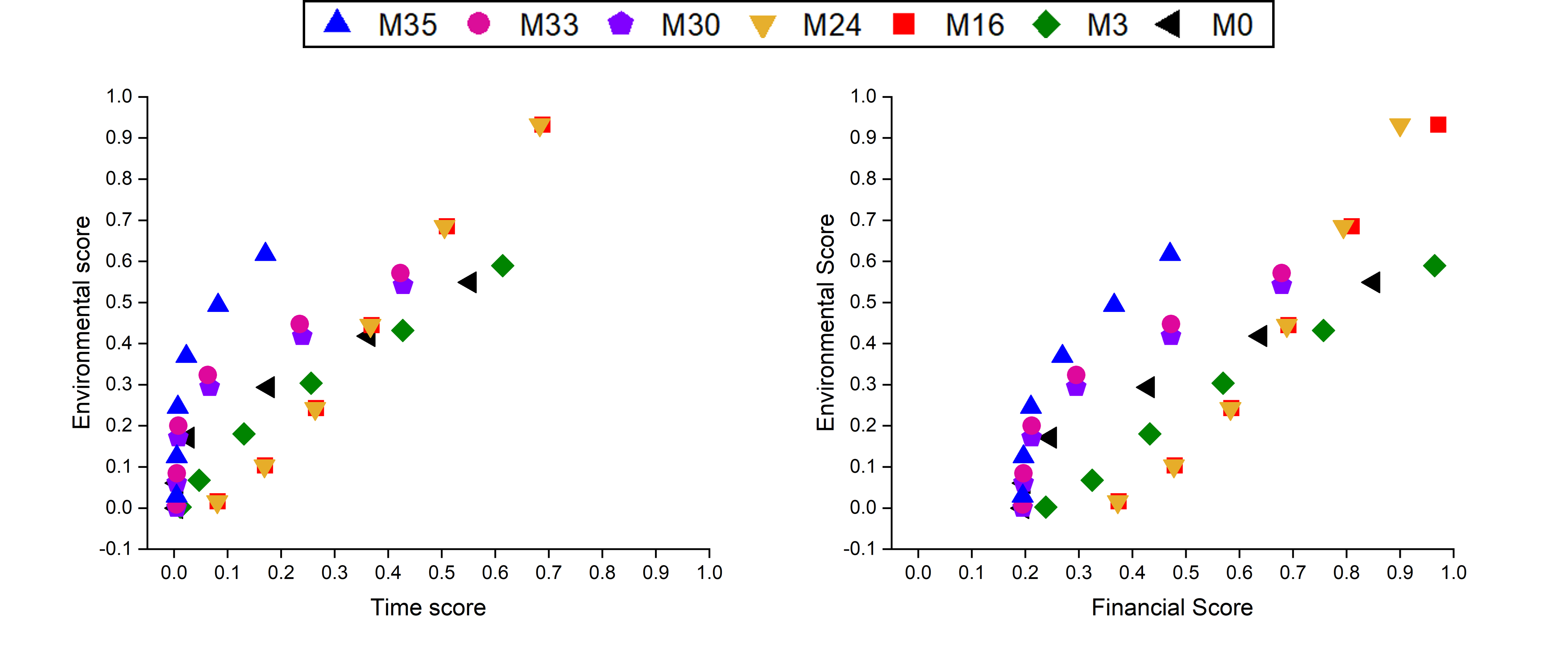}
         \caption{
    Single Day Impact by Varying the Volume ($\#doc$) on Data Models' Time and Environmental and Financial costs, with all the Queries ($\#srv = 1,000 $)}
         \label{fig:datavariation}
\end{figure*}

\subsubsection{Query's Cost}

In order to study how distinct data model costs vary on different types of queries, we applied the 5 queries in Table~\ref{tab:queries}. For each of these queries, we extract used keys and joins to study their impact on distinct structures of data models.

Figure~\ref{fig:queriescost} depicts the time \textit{vs.} environmental costs in logarithmic scales on a setting of 1M warehouses and 1k servers.
These query costs are given for a single occurrence simulation. The dashed red lines represent the time constraint ($\tau_{t_q}$) for each query which corresponds to query's max execution time defined by decision makers. Data models at the left side of upper-bound lines are the qualified data models $M^A$.

We can notice that $M0$ (black triangles) minimizes all filter queries' cost since they always target one row and have a low impact on memory reads (documents are shorter). Thus, for each query only one row is processed. When applied to join queries it costs more with such a data model since it requires supplementary steps.

Data model $M35$ (blue triangles) is a fully denormalized data model that consists of only one row $O$ where $C$ and $W$ are merged. Merges help to avoid costly joins, and we notice how $M35$ minimizes the join queries especially $\mathcal{Q}5$ that requires all keys, which are all in a single row in this data model.
It also provides  a good trade-off on filter queries.

If we compare $M35$ with $M16$ (red squares) and $M24$ (yellow triangles), which also consist of merges of $O$ and $W$ in $C$,  they show the highest costs for all queries.
This is due to the cardinality problem, since each customer can have multiple orders thus, the size of row $C$ becomes huge and requires more memory reads. Moreover, they do not contain all necessary keys which implies joins for $\mathcal{Q}4$ and $\mathcal{Q}5$.
Compared to $M35$, each order corresponds to one customer and one warehouse thus, $O$ contains all keys but still does not cost as much as others.

Moreover, split data models like $M16$, $M24$ and $M3$ (green diamonds) require document reconstructions for some filter queries as shown is $\mathcal{Q}2$ and $\mathcal{Q}1$. This document reconstruction must be done by a supplementary join which is not required for other data models. We can see that the split on $M3$ has a positive impact for $\mathcal{Q}2$ but not on $\mathcal{Q}1$.

Last, data models $M30$ (purple pentagons) and $M33$ (pink circles) which apply a simple merge show a good trade-off on most queries except for $\mathcal{Q}5$ which join has not been optimized (like in $M35$). However, this query has a low occurrence and a few global impact.

\subsubsection{Total Data Model's Cost}

In order to study the global behavior of data models, we have calculated costs of data models on different combinations of settings (\textit{i.e.,} data volume and number of servers).

For our simulation, and in order to study the impact of data volume, we fix the number of servers to $\#srv = 1,000$ and we vary data volumes from 1k to 100M warehouses while factoring by 10.
Figure~\ref{fig:datavariation} depicts the total time and financial scores wrt. to environmental scores.
Scores are normalized costs 
for every data volume variation of the same 7 data models proposed.  

\paragraph{\textit{Time costs} (left chart)}
We notice that $M16$ and $M24$ grow exponentially in terms of both time and environmental scores which is due to large-size rows. Both data models consist of a row $C$ with nested $O$ and $W$. This row's size grows noticeably with the increase in data volume thus, operations on these rows become very costly.

Likewise other data models witness a stronger growth of environmental scores compared to time until reaching 100k warehouses.
In fact, the data model's size growth impacts storage cost and thus, environmental cost. While over 100k warehouses, with more data the processing and transferring times become costlier.

It is interesting to notice that $\mathcal{M}30$ and $\mathcal{M}33$ witness a fast growth up to 100k warehouses on environmental impact  while it slows down more than other solutions afterwards. The merge combination between $O$ and $C$ helps to reduce important joins while not impacting too much filter queries on huge documents (high number of warehouses).

\paragraph{\textit{Financial scores} (right chart)}
Small differences can be shown in these settings with the volume of data transferred externally (servers' cost is constant). 
Same as for time score, we notice that $M16$ and $M24$ grow the most on both financial and environmental dimensions. These two data models contain multiple splits that are not required for all queries, thus they generate more transfers mostly for the join queries\footnote{Some join queries can be done at cluster side if the NoSQL solution allows it, it will reduce the financial cost but not the time cost (same operation).}. This transfer cost impacts both environmental and financial costs.

For the rest of data models, when data is relatively small (up to 100k warehouses), we notice more impact on the environmental score since it integrates processing, storage and transfer costs while financial cost is only impacted by external transfer cost. Therefore, only when data volume becomes bigger than transfer cost becomes noticeable.

We must notice that normalized financial scores begin around 0.2 since it relies mostly on the size of the cluster (number of servers). Here, we fixed it to 1,000 servers and the global cost will be lower with less servers (10 ones). For data visualization purpose, we did not show all settings.

\section{Conclusion}
\label{sec:conclusion}

In this paper, we have proposed a multidimensional cost model that allows  choosing the optimal data model out of different possible solutions thanks to a stability measure of data models.
Those solutions generated by denormalization are compatible with a given use case.

In order to compare data models and determine the optimal one(s), our multidimensional cost model integrates time, environmental and financial dimensions to define the cost of each data model. It considers queries cost on a data model and cost of data model itself. The costs depend on volume of data processed, transferred and stored thus, it is impacted by changes of settings (\textit{i.e.,} data volume, number of servers).

For future works, we are designing a demonstrator to ease the decision maker to understand the impact of parameters and data models.
We would like to complete our cost model and analysis with aggregate queries' costs.
In fact, NoSQL databases rely on the Map/Reduce to process aggregate queries, our cost model must take into account each phase \textit{map} (read \& emit), \textit{shuffle} (intra-datacenter communications) and \textit{reduce} (local and global group by). Even if it does not change our approach, the analysis will be more complete. 
Moreover, we would like to study thoroughly the impact of denormalizations on data models costs. Currently, each type of denormalization impacts the cost of a certain type of queries. This study will help to improve data models generation process by avoiding applying denormalizations that are more costly.



\bibliographystyle{ACM-Reference-Format}
\bibliography{references}


\begin{thebibliography}{25}


\ifx \showCODEN    \undefined \def \showCODEN     #1{\unskip}     \fi
\ifx \showDOI      \undefined \def \showDOI       #1{#1}\fi
\ifx \showISBNx    \undefined \def \showISBNx     #1{\unskip}     \fi
\ifx \showISBNxiii \undefined \def \showISBNxiii  #1{\unskip}     \fi
\ifx \showISSN     \undefined \def \showISSN      #1{\unskip}     \fi
\ifx \showLCCN     \undefined \def \showLCCN      #1{\unskip}     \fi
\ifx \shownote     \undefined \def \shownote      #1{#1}          \fi
\ifx \showarticletitle \undefined \def \showarticletitle #1{#1}   \fi
\ifx \showURL      \undefined \def \showURL       {\relax}        \fi
\providecommand\bibfield[2]{#2}
\providecommand\bibinfo[2]{#2}
\providecommand\natexlab[1]{#1}
\providecommand\showeprint[2][]{arXiv:#2}

\bibitem[Abdelhedi et~al\mbox{.}(2017)]%
        {abdelhedi2017}
\bibfield{author}{\bibinfo{person}{Fatma Abdelhedi}, \bibinfo{person}{Amal~Ait
  Brahim}, \bibinfo{person}{Faten Atigui}, {and} \bibinfo{person}{Gilles
  Zurfluh}.} \bibinfo{year}{2017}\natexlab{}.
\newblock \showarticletitle{{MDA-based Approach for NoSQL Databases
  Modelling}}. In \bibinfo{booktitle}{\emph{DAWAK'17}}. Springer,
  \bibinfo{pages}{88--102}.
\newblock


\bibitem[Afrati and Ullman(2010)]%
        {10.1145/1739041.1739056}
\bibfield{author}{\bibinfo{person}{Foto~N. Afrati} {and}
  \bibinfo{person}{Jeffrey~D. Ullman}.} \bibinfo{year}{2010}\natexlab{}.
\newblock \showarticletitle{Optimizing Joins in a Map-Reduce Environment}. In
  \bibinfo{booktitle}{\emph{Proceedings of the 13th International Conference on
  Extending Database Technology}} (Lausanne, Switzerland)
  \emph{(\bibinfo{series}{EDBT '10})}. \bibinfo{publisher}{Association for
  Computing Machinery}, \bibinfo{address}{New York, NY, USA},
  \bibinfo{pages}{99–110}.
\newblock
\showISBNx{9781605589459}
\urldef\tempurl%
\url{https://doi.org/10.1145/1739041.1739056}
\showDOI{\tempurl}


\bibitem[Al-Badarneh and Rababa(2022)]%
        {ALBADARNEH20221074}
\bibfield{author}{\bibinfo{person}{Amer~F. Al-Badarneh} {and}
  \bibinfo{person}{Salahaldeen~Atef Rababa}.} \bibinfo{year}{2022}\natexlab{}.
\newblock \showarticletitle{An analysis of two-way equi-join algorithms under
  MapReduce}.
\newblock \bibinfo{journal}{\emph{Journal of King Saud University - Computer
  and Information Sciences}} \bibinfo{volume}{34}, \bibinfo{number}{4}
  (\bibinfo{year}{2022}), \bibinfo{pages}{1074--1085}.
\newblock
\showISSN{1319-1578}
\urldef\tempurl%
\url{https://doi.org/10.1016/j.jksuci.2020.05.004}
\showDOI{\tempurl}


\bibitem[Blanas et~al\mbox{.}(2010)]%
        {10.1145/1807167.1807273}
\bibfield{author}{\bibinfo{person}{Spyros Blanas}, \bibinfo{person}{Jignesh~M.
  Patel}, \bibinfo{person}{Vuk Ercegovac}, \bibinfo{person}{Jun Rao},
  \bibinfo{person}{Eugene~J. Shekita}, {and} \bibinfo{person}{Yuanyuan Tian}.}
  \bibinfo{year}{2010}\natexlab{}.
\newblock \showarticletitle{A Comparison of Join Algorithms for Log Processing
  in MaPreduce}. In \bibinfo{booktitle}{\emph{Proceedings of the 2010 ACM
  SIGMOD International Conference on Management of Data}} (Indianapolis,
  Indiana, USA) \emph{(\bibinfo{series}{SIGMOD '10})}.
  \bibinfo{publisher}{Association for Computing Machinery},
  \bibinfo{address}{New York, NY, USA}, \bibinfo{pages}{975–986}.
\newblock
\showISBNx{9781450300322}
\urldef\tempurl%
\url{https://doi.org/10.1145/1807167.1807273}
\showDOI{\tempurl}


\bibitem[Chebotko et~al\mbox{.}(2015)]%
        {chebotko2015}
\bibfield{author}{\bibinfo{person}{Artem Chebotko}, \bibinfo{person}{Andrey
  Kashlev}, {and} \bibinfo{person}{Shiyong Lu}.}
  \bibinfo{year}{2015}\natexlab{}.
\newblock \showarticletitle{{A Big Data Modeling Methodology for Apache
  Cassandra}}. In \bibinfo{booktitle}{\emph{ICBD'15}}. IEEE,
  \bibinfo{pages}{238--245}.
\newblock


\bibitem[Daniel et~al\mbox{.}(2016)]%
        {daniel2016umltographdb}
\bibfield{author}{\bibinfo{person}{Gwendal Daniel}, \bibinfo{person}{Gerson
  Suny{\'e}}, {and} \bibinfo{person}{Jordi Cabot}.}
  \bibinfo{year}{2016}\natexlab{}.
\newblock \showarticletitle{{UMLtoGraphDB: mapping conceptual schemas to graph
  databases}}. In \bibinfo{booktitle}{\emph{Conceptual Modeling (ER'16)}}.
  \bibinfo{pages}{430--444}.
\newblock


\bibitem[de~Freitas et~al\mbox{.}(2016)]%
        {claudino2016}
\bibfield{author}{\bibinfo{person}{Myller~Claudino de Freitas},
  \bibinfo{person}{Damires~Yluska Souza}, {and} \bibinfo{person}{Ana Salgado}.}
  \bibinfo{year}{2016}\natexlab{}.
\newblock \showarticletitle{Conceptual Mappings to Convert Relational into
  NoSQL Databases}. In \bibinfo{booktitle}{\emph{ICEIS'16}}.
  \bibinfo{pages}{174--181}.
\newblock


\bibitem[{de la Vega} et~al\mbox{.}(2020)]%
        {DELAVEGA2020455}
\bibfield{author}{\bibinfo{person}{Alfonso {de la Vega}},
  \bibinfo{person}{Diego Garc\'ia-Saiz}, \bibinfo{person}{Carlos Blanco},
  \bibinfo{person}{Marta Zorrilla}, {and} \bibinfo{person}{Pablo S\'anchez}.}
  \bibinfo{year}{2020}\natexlab{}.
\newblock \showarticletitle{Mortadelo: Automatic generation of NoSQL stores
  from platform-independent data models}.
\newblock \bibinfo{journal}{\emph{Future Generation Computer Systems}}
  \bibinfo{volume}{105} (\bibinfo{year}{2020}), \bibinfo{pages}{455--474}.
\newblock
\showISSN{0167-739X}


\bibitem[Dean and Ghemawat(2004)]%
        {10.5555/1251254.1251264}
\bibfield{author}{\bibinfo{person}{Jeffrey Dean} {and} \bibinfo{person}{Sanjay
  Ghemawat}.} \bibinfo{year}{2004}\natexlab{}.
\newblock \showarticletitle{MapReduce: Simplified Data Processing on Large
  Clusters} \emph{(\bibinfo{series}{OSDI'04})}. \bibinfo{publisher}{USENIX
  Association}, \bibinfo{address}{USA}, \bibinfo{pages}{10}.
\newblock


\bibitem[Gupta et~al\mbox{.}(2022)]%
        {10.1109/MM.2022.3163226}
\bibfield{author}{\bibinfo{person}{Udit Gupta}, \bibinfo{person}{Young~Geun
  Kim}, \bibinfo{person}{Sylvia Lee}, \bibinfo{person}{Jordan Tse},
  \bibinfo{person}{Hsien-Hsin~S. Lee}, \bibinfo{person}{Gu-Yeon Wei},
  \bibinfo{person}{David Brooks}, {and} \bibinfo{person}{Carole-Jean Wu}.}
  \bibinfo{year}{2022}\natexlab{}.
\newblock \showarticletitle{Chasing Carbon: The Elusive Environmental Footprint
  of Computing}.
\newblock \bibinfo{journal}{\emph{IEEE Micro}} \bibinfo{volume}{42},
  \bibinfo{number}{4} (\bibinfo{date}{jul} \bibinfo{year}{2022}),
  \bibinfo{pages}{37–47}.
\newblock
\showISSN{0272-1732}
\urldef\tempurl%
\url{https://doi.org/10.1109/MM.2022.3163226}
\showDOI{\tempurl}


\bibitem[Guyon(2018)]%
        {guyon2018}
\bibfield{author}{\bibinfo{person}{David Guyon}.}
  \bibinfo{year}{2018}\natexlab{}.
\newblock \emph{\bibinfo{title}{Supporting energy-awareness for cloud users}}.
\newblock \bibinfo{thesistype}{Ph.\,D. Dissertation}.
\newblock
\urldef\tempurl%
\url{http://www.theses.fr/2018REN1S037/document}
\showURL{%
\tempurl}
\newblock
\shownote{2018REN1S037}.


\bibitem[Li et~al\mbox{.}(2014)]%
        {li2014}
\bibfield{author}{\bibinfo{person}{Yan Li}, \bibinfo{person}{Ping Gu}, {and}
  \bibinfo{person}{Chao Zhang}.} \bibinfo{year}{2014}\natexlab{}.
\newblock \showarticletitle{{Transforming UML class diagrams into HBase based
  on meta-model}}. In \bibinfo{booktitle}{\emph{{ISEEE'14}}},
  Vol.~\bibinfo{volume}{2}. \bibinfo{pages}{720--724}.
\newblock


\bibitem[Mahajan et~al\mbox{.}(2019)]%
        {mahajan2019}
\bibfield{author}{\bibinfo{person}{Divya Mahajan}, \bibinfo{person}{Cody
  Blakeney}, {and} \bibinfo{person}{Ziliang Zong}.}
  \bibinfo{year}{2019}\natexlab{}.
\newblock \showarticletitle{Improving the energy efficiency of relational and
  NoSQL databases via query optimizations}.
\newblock \bibinfo{journal}{\emph{Sustainable Computing: Informatics and
  Systems}}  \bibinfo{volume}{22} (\bibinfo{year}{2019}),
  \bibinfo{pages}{120--133}.
\newblock


\bibitem[Mali et~al\mbox{.}(2022)]%
        {Mali2022}
\bibfield{author}{\bibinfo{person}{Jihane Mali}, \bibinfo{person}{Shohreh
  Ahvar}, \bibinfo{person}{Faten Atigui}, \bibinfo{person}{Ahmed Azough}, {and}
  \bibinfo{person}{Nicolas Travers}.} \bibinfo{year}{2022}\natexlab{}.
\newblock \showarticletitle{A Global Model-Driven Denormalization Approach for
  Schema Migration}. In \bibinfo{booktitle}{\emph{International Conference on
  Research Challenges in Information Science}}. Springer,
  \bibinfo{pages}{529--545}.
\newblock


\bibitem[Mali et~al\mbox{.}(2020)]%
        {Mali2020}
\bibfield{author}{\bibinfo{person}{Jihane Mali}, \bibinfo{person}{Faten
  Atigui}, \bibinfo{person}{Ahmed Azough}, {and} \bibinfo{person}{Nicolas
  Travers}.} \bibinfo{year}{2020}\natexlab{}.
\newblock \showarticletitle{{ModelDrivenGuide: An Approach for Implementing
  NoSQL Schemas}}. In \bibinfo{booktitle}{\emph{DEXA'20}}. Springer,
  \bibinfo{pages}{141--151}.
\newblock


\bibitem[{\"O}zsu and Valduriez(1999)]%
        {ozsu1999}
\bibfield{author}{\bibinfo{person}{M~Tamer {\"O}zsu} {and}
  \bibinfo{person}{Patrick Valduriez}.} \bibinfo{year}{1999}\natexlab{}.
\newblock \bibinfo{booktitle}{\emph{Principles of distributed database
  systems}}. Vol.~\bibinfo{volume}{2}.
\newblock \bibinfo{publisher}{Springer}.
\newblock


\bibitem[Rocha et~al\mbox{.}(2015)]%
        {rocha2015}
\bibfield{author}{\bibinfo{person}{Leonardo Rocha}, \bibinfo{person}{Fernando
  Vale}, \bibinfo{person}{Elder Cirilo}, \bibinfo{person}{D{\'a}rlinton
  Barbosa}, {and} \bibinfo{person}{Fernando Mour{\~a}o}.}
  \bibinfo{year}{2015}\natexlab{}.
\newblock \showarticletitle{{A framework for migrating relational datasets to
  NoSQL}}.
\newblock \bibinfo{journal}{\emph{Procedia Computer Science}}
  \bibinfo{volume}{51} (\bibinfo{year}{2015}), \bibinfo{pages}{2593--2602}.
\newblock


\bibitem[Rodriguez-Martinez et~al\mbox{.}(2011)]%
        {rodriguez2011}
\bibfield{author}{\bibinfo{person}{Manuel Rodriguez-Martinez},
  \bibinfo{person}{Harold Valdivia}, \bibinfo{person}{Jaime Seguel}, {and}
  \bibinfo{person}{Melvin Greer}.} \bibinfo{year}{2011}\natexlab{}.
\newblock \showarticletitle{Estimating power/energy consumption in database
  servers}.
\newblock \bibinfo{journal}{\emph{Procedia Computer Science}}
  \bibinfo{volume}{6} (\bibinfo{year}{2011}), \bibinfo{pages}{112--117}.
\newblock


\bibitem[Saraiva et~al\mbox{.}(2017)]%
        {saraiva2017}
\bibfield{author}{\bibinfo{person}{Jo{\~a}o Saraiva}, \bibinfo{person}{Miguel
  Guimarales}, {and} \bibinfo{person}{Orlando Belo}.}
  \bibinfo{year}{2017}\natexlab{}.
\newblock \showarticletitle{An economic energy approach for queries on data
  centers}.
\newblock  (\bibinfo{year}{2017}).
\newblock


\bibitem[Sellami and Defude(2017)]%
        {sellami2017}
\bibfield{author}{\bibinfo{person}{Rami Sellami} {and} \bibinfo{person}{Bruno
  Defude}.} \bibinfo{year}{2017}\natexlab{}.
\newblock \showarticletitle{Complex queries optimization and evaluation over
  relational and NoSQL data stores in cloud environments}.
\newblock \bibinfo{journal}{\emph{IEEE transactions on big data}}
  \bibinfo{volume}{4}, \bibinfo{number}{2} (\bibinfo{year}{2017}),
  \bibinfo{pages}{217--230}.
\newblock


\bibitem[Smith(1985)]%
        {10.1145/3959.3961}
\bibfield{author}{\bibinfo{person}{Alan~J. Smith}.}
  \bibinfo{year}{1985}\natexlab{}.
\newblock \showarticletitle{Disk Cache—miss Ratio Analysis and Design
  Considerations}.
\newblock \bibinfo{journal}{\emph{ACM Trans. Comput. Syst.}}
  \bibinfo{volume}{3}, \bibinfo{number}{3} (\bibinfo{date}{aug}
  \bibinfo{year}{1985}), \bibinfo{pages}{161–203}.
\newblock
\showISSN{0734-2071}
\urldef\tempurl%
\url{https://doi.org/10.1145/3959.3961}
\showDOI{\tempurl}


\bibitem[Tannu and Nair(2022)]%
        {tannu2022dirty}
\bibfield{author}{\bibinfo{person}{Swamit Tannu} {and}
  \bibinfo{person}{Prashant~J. Nair}.} \bibinfo{year}{2022}\natexlab{}.
\newblock \bibinfo{title}{The Dirty Secret of SSDs: Embodied Carbon}.
\newblock
\newblock
\showeprint[arxiv]{2207.10793}~[cs.AR]


\bibitem[Tauro et~al\mbox{.}(2012)]%
        {tauro2012}
\bibfield{author}{\bibinfo{person}{Clarence~JM Tauro},
  \bibinfo{person}{Shreeharsha Aravindh}, {and} \bibinfo{person}{AB
  Shreeharsha}.} \bibinfo{year}{2012}\natexlab{}.
\newblock \showarticletitle{{Comparative study of the new generation, agile,
  scalable, high performance NoSQL databases}}.
\newblock \bibinfo{journal}{\emph{{International Journal of Computer
  Applications}}} \bibinfo{volume}{48}, \bibinfo{number}{20}
  (\bibinfo{year}{2012}), \bibinfo{pages}{1--4}.
\newblock


\bibitem[Thakur and Chaurasia(2016)]%
        {thakur2016carbon}
\bibfield{author}{\bibinfo{person}{Sanjeev Thakur} {and} \bibinfo{person}{Ankur
  Chaurasia}.} \bibinfo{year}{2016}\natexlab{}.
\newblock \showarticletitle{Towards Green Cloud Computing: Impact of carbon
  footprint on environment}. In \bibinfo{booktitle}{\emph{2016 6th
  international conference-cloud system and big data engineering
  (Confluence)}}. IEEE, \bibinfo{pages}{209--213}.
\newblock


\bibitem[Vajk et~al\mbox{.}(2013)]%
        {vajk2013}
\bibfield{author}{\bibinfo{person}{Tam{\'a}s Vajk}, \bibinfo{person}{P{\'e}ter
  Feh{\'e}r}, \bibinfo{person}{Kriszti{\'a}n Fekete}, {and}
  \bibinfo{person}{Hassan Charaf}.} \bibinfo{year}{2013}\natexlab{}.
\newblock \showarticletitle{Denormalizing data into schema-free databases}. In
  \bibinfo{booktitle}{\emph{{CogInfoCom'13}}}. IEEE, \bibinfo{pages}{747--752}.
\newblock


\end{thebibliography}

\end{document}